\documentclass[%
 reprint,
 amsmath,amssymb,
 aps,
 prl,
]{revtex4-2}

\usepackage{comment}
\usepackage[]{graphicx}
\usepackage[caption=false]{subfig}
\usepackage{dblfloatfix}
\usepackage{natbib}
\usepackage{amsmath}
\usepackage[version=4]{mhchem}
\usepackage{placeins}
\usepackage{physics}
\usepackage{siunitx}
\usepackage{dcolumn}
\usepackage{bm}
\usepackage[hyperfootnotes=false]{hyperref}
\usepackage{array}
\newcolumntype{P}[1]{>{\centering\arraybackslash}p{#1}}
\hypersetup{colorlinks=true,
        allcolors = blue,
		}

\begin{document}
\preprint{APS/123-QED}

\title{Experimental demonstration of a versatile and scalable scheme for iterative generation of non-Gaussian states of light}

\author{Hector Simon, Lucas Caron, Romaric Journet, Viviane Cotte and Rosa Tualle-Brouri}%
 \email{rosa.tualle-brouri@institutoptique.fr}
\affiliation{%
 Laboratoire Charles Fabry, Institut d'Optique Graduate School, Université Paris-Saclay.\\
}%

\date{\today}

\begin{abstract}

Non-Gaussian states of light, such as GKP states, are essential resources for optical continuous-variable quantum computing. The ability to efficiently produce these states would open up tremendous prospects for quantum technologies in general and fault-tolerant quantum computing in particular.
This letter demonstrates a versatile method using a quantum memory cavity to overcome the probabilistic nature of the breeding protocols and generate non-Gaussian states at high rates with scalability perspectives.
The performances of our experimental setup are illustrated with the generation of Schrödinger cat states of amplitude $\alpha=1.63$ with a fidelity of more than \SI{60}{\%} at a generation rate in the \SI{}{kHz} range, which is higher than the state of the art for such states.

\end{abstract}

\maketitle

\section{Introduction}

Breeding operations \cite{lund_conditional_2004,suzuki_practical_2006,etesse_experimental_2015}, which consist in the conditional generation of complex quantum states from smaller resources, are the subject of growing interest. Applied iteratively, they could allow the generation of GKP \cite{gottesman_encoding_2001} states on traveling light pulses \cite{vasconcelos_all-optical_2010,etesse_proposal_2014,Eaton_2019}, a cornerstone in the field of all-optical quantum computing \cite{gottesman_encoding_2001,baragiola_all-gaussian_2019, lvovsky_production_2020}. Many works focused on the simplest breeding operation, starting from two single photons \cite{etesse_experimental_2015} or two squeezed single photons (kitten states) \cite{sychev_enlargement_2017,konno_logical_2024} and mixing them on a symmetric beam-splitter. The breeding operation is then heralded by a quadrature measurement on one output port of the beam-splitter. This leads to a quantum state with a richer structure that was recently mentioned as a faint GKP state \cite{konno_logical_2024}, even if it is admitted that those states are still far from beeing suitable for quantum computing \cite{kawasaki_high-rate_2024}. 
It is worth noticing that the states presented in all these works are very similar, up to a squeezing operation. In the present work, we will refer to it as a squeezed Schrödinger cat state. 

Iteratively applying breeding operations could open new perspectives, but their probabilistic nature raises a major issue. The success of the whole protocol, requires the success of each intermediate step, including the conditional generation of the initial resources. Therefore, the probabilistic nature of the generation protocols and the rapidly growing number of required resource states make the practical implementation of the experiment extremely challenging, and its success probability 
inconveniently low.

\begin{figure}[h!]
\subfloat[\label{fig:mult_temp}]{\includegraphics[scale=0.8]{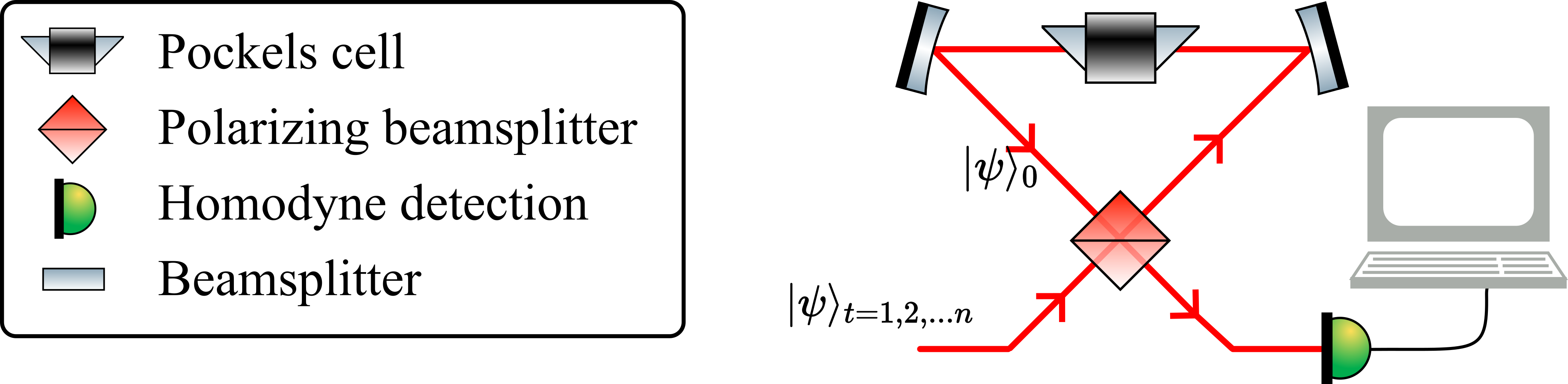}}\\
\subfloat[\label{fig:mult_spat}]{\includegraphics[scale=0.8]{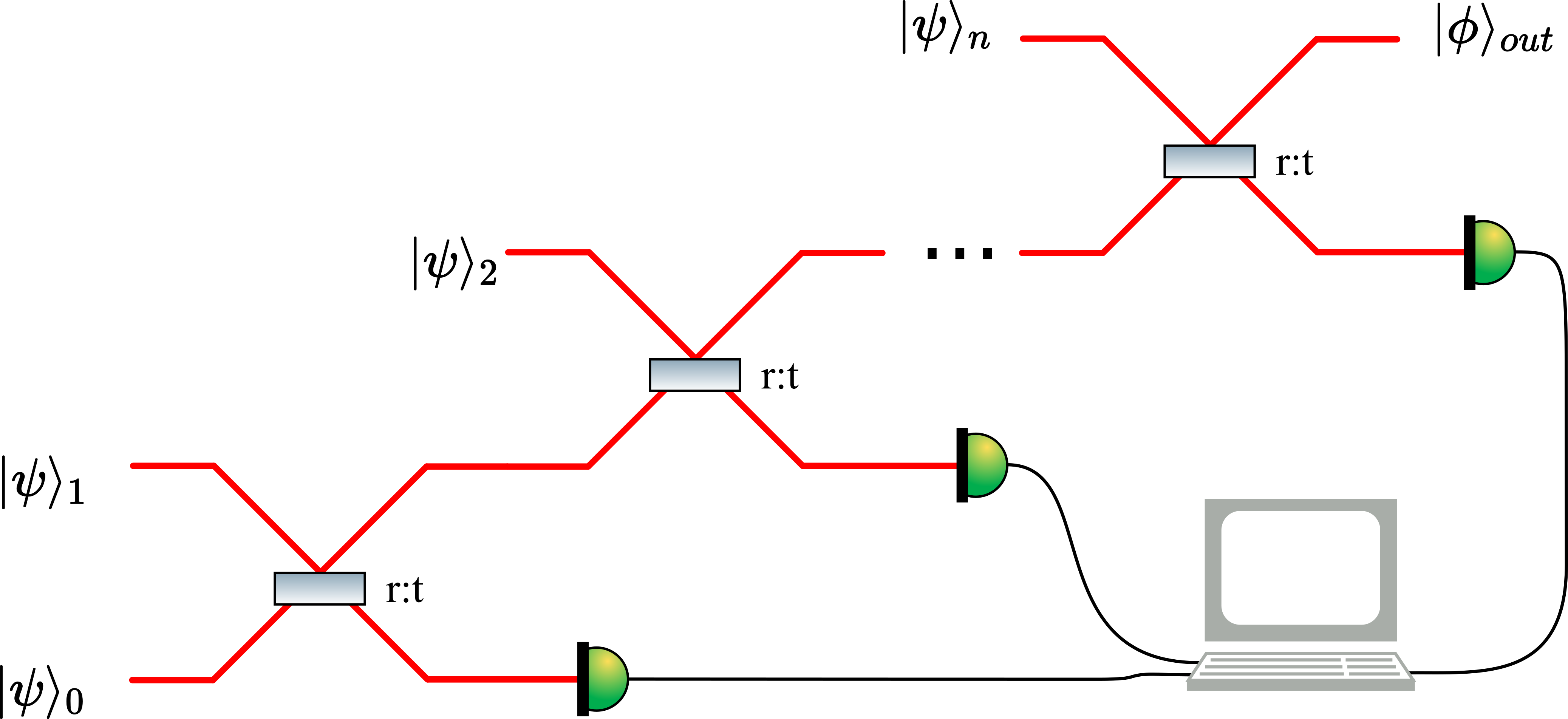}}
\caption{\label{fig:mult_spat_temp} \normalsize{Schematic representation of feed forward generation method using a quantum-memory cavity (\ref{fig:mult_temp}). Equivalent scheme using a beam-splitter network (\ref{fig:mult_spat}).}}
\end{figure}

We propose a different approach to 
meet this scalability challenge, based on active temporal multiplexing, using a quantum memory cavity \cite{cotte_experimental_2022,bouillard_quantum_2019}. In this case, a reduced number of beam-splitters is required and the intermediate resource states of the  protocol do not need to be generated
simultaneously for the entire protocol to succeed.\\
The setup presented in this letter is a compact solution to efficiently implement protocols that consist in the growth of a non-Gaussian state using successive breeding steps with a non-deterministically generated resource state. In the proposed scheme (Fig. \ref{fig:mult_temp}),
a picosecond optical pulse periodically interacts with a beam-splitter with rapidly adjustable reflectivity. At each interaction, we can decide whether the beam-splitter will behave as a node in the equivalent network (Fig. \ref{fig:mult_spat}) or if it will behave as an edge by tuning its reflectivity to a value close to \SI{100}{\%}. 
This method allows us to be sure that a resource state is available before creating a node, leading to a significant increase in the success probability of an iterative generation protocol. In this paper we illustrate the performances of this setup with the generation of even Schrödinger cat states (SCS) at a rate higher than \SI{1}{kHz}, which is about one order of magnitude beyond the state of the art for such states \cite{huang_optical_2015} with a fidelity of more than \SI{60}{\%} to an ideal squeezed SCS of amplitude $\alpha=1.63$.

\section{\label{sec:level1}Schrödinger cat state generation}

The protocol implemented on the aforementioned experimental setup for the SCS generation consists in two steps \cite{etesse_experimental_2015}: the generation of a two-photons NOON state through the coalescence of two photons on a balanced beam-splitter, and a $\hat{X}$ quadrature measurement on one output port of the beam-splitter. If the measurement result is sufficiently close to $X=0$, the state in the other port is projected on the desired state $\ket{\psi}_{cat}$, which has a $99\%$ fidelity to a squeezed coherent state superposition of amplitude $\alpha=1.63$ squeezed of $\SI{3.64}{dB}$  along the $\hat{X}$ direction.

\subsection{\label{sec:level2}Experimental setup}

\begin{figure}[h!]
\includegraphics[scale=0.7]{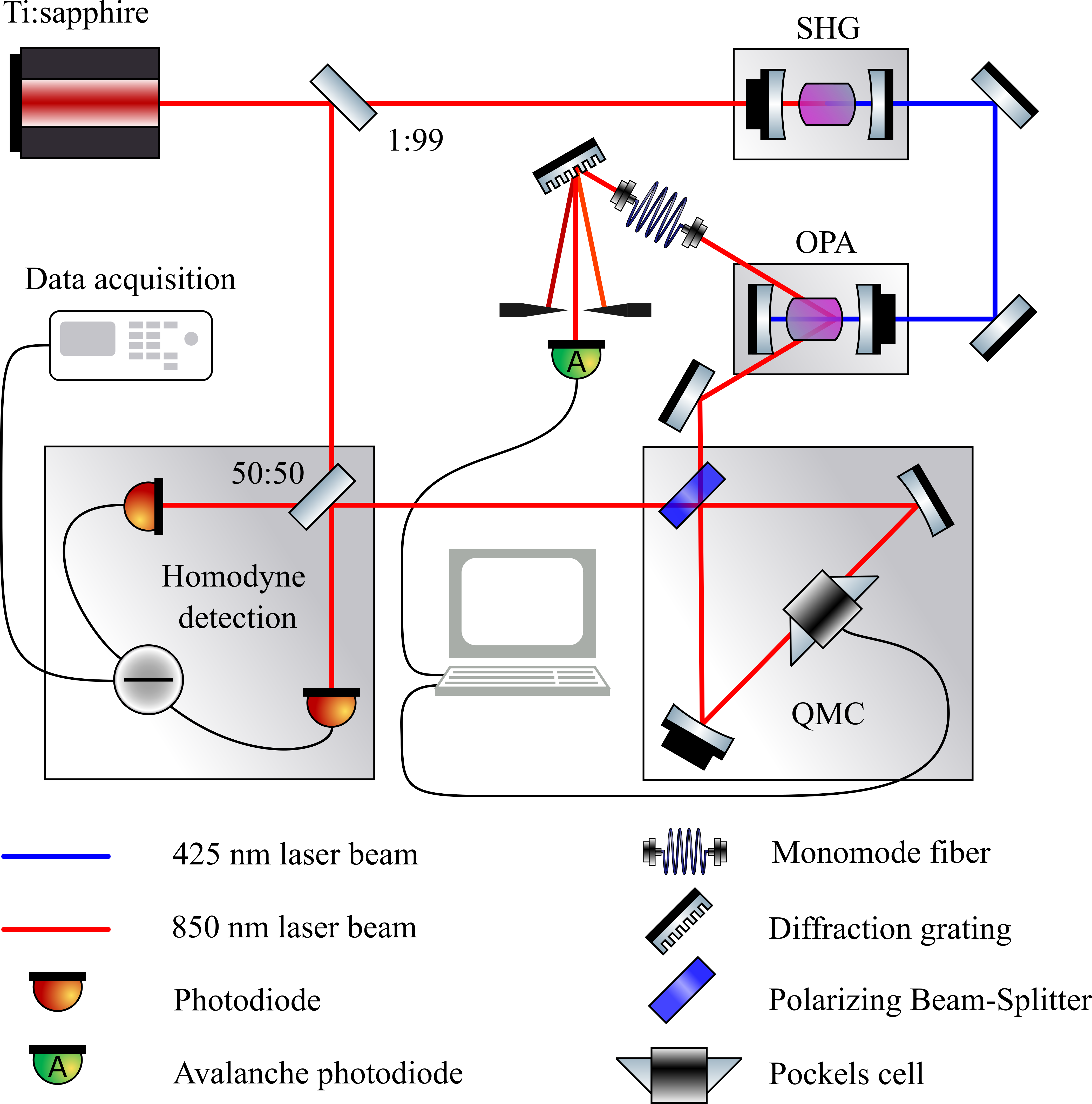}
\caption{\label{fig:schema_manip}\normalsize{Simplified experimental setup with the single photon source, the quantum-memory cavity, and the homodyne detection.}}
\end{figure}

The basic resource states in this experiment, the single photons, are generated by spontaneous parametric down-conversion \cite{hong_experimental_1986} using a source previously described in \cite{bouillard_high_2019}. The laser source of our experiment (Fig. \ref{fig:schema_manip}) is a pulsed (\SI{76}{MHz}) Ti-sapphire at \SI{850}{nm}. The pulse duration is \SI{2.2}{ps} and the average power is \SI{390}{mW}. The single-photon source consists of two optical cavities, each including a non-linear crystal (\ce{BiB3O6}), which increase the desired non-linear effects by enhancing the pump beam. The first cavity constitutes a second harmonic generation stage (SHG) \cite{kanseri_efficient_2016} and the second one is an optical parametric amplifier (OPA).
The SHG is pumped at \SI{280}{mW} average power and produces frequency-doubled pulses with an average power of up to \SI{150}{mW} at \SI{425}{nm}. 
The intracavity average power in the OPA can reach a maximum of \SI{6}{W}. The OPA uses a non-colinear type-I phase matching to produce a two-mode squeezed vacuum state. 
The signal beam of this state is coupled to a monomode fiber and spectrally filtered with a diffraction grating and a slit before being coupled to an avalanche photodiode (APD) with a \SI{45}{\%} detection efficiency. A detection event on the APD heralds a photon in a well-defined mode in the idler beam. The heralding rate of this source is typically \SI{300}{kHz} and the single photon fidelity is $F \simeq 87\%$, corrected for the 76\% detection efficiency of the homodyne detection used for its tomography (see Supplemental Material \cite{sm}). We use a \SI{60}{m}-long Herriott-cell-like delay line to introduce a \SI{200}{ns} delay between the APD click and the single photon's arrival in the next stage of the experiment. This delay allows us to drive the experimental setup in real time after each heralding event.

The main originality of this setup is a quantum-memory cavity (QMC) that can be used for both quantum states' storage and breeding operations \cite{cotte_experimental_2022,bouillard_quantum_2019}. This device consists of a low-loss optical cavity containing a fast Pockels cell (PC) and a polarizing beam-splitter. The cavity length matches the repetition rate of the laser. This setup is equivalent to an optical cavity closed with a tunable beam-splitter. We tune the reflectivity of the beam-splitter via polarization control using the PC inside the cavity. 
The heralded single photon is transmitted through the polarizing beam-splitter, thus entering the QMC. We apply a $V_\pi$ voltage to the PC for a duration shorter than one round trip of the cavity. The photon's polarization is switched when passing through the PC, trapping the photon inside the QMC until a non-zero voltage is applied to the PC. 
After a second photon is heralded, we apply a $V_{\pi/2}$ voltage when the photons arrive on the PC, which switch their polarization to right and left circular polarizations \cite{cotte_experimental_2022}. Both photons then experience the polarizing beam-splitter as a balanced beam-splitter. 

The result of this coalescence operation is a two-photons NOON state.
The photons in each mode of this state are polarized in one of the eigenmodes of the polarizing beam-splitter. Using a homodyne detection, we perform a quadrature measurement on the state that leaves the QMC. For a measurement result of $\hat{X}$ close to $X=0$, we generate the state $\ket{\psi}_{\text{cat}}$ in the QMC. In this experiment, we chose $X\in[-0.3,0.3]$, which corresponds to a conditioning probability of $23.5\%$. The SCS is stored in the QMC and can be extracted for characterization or to be used for a more complex protocol.

In the present case, we are only interested in characterizing the generated SCS. To do so, we use the same homodyne detection
we used for the conditioning measurement. Due to the bandwidth of the homodyne detection (10 MHz), we store the SCS for 15 round trips before applying a $V_{\pi}$ voltage to the PC to extract and measure it. 
This state is however directly  available as soon as it was created, it is therefore relevant to characterize it at this stage, before it undergoes storage losses.

Due to the phase dependence of the Wigner function of the generated state, we need to measure the phase of the generated state relative to the local oscillator. This measurement is done at each SCS generation attempt. The measurement protocol is detailed in the Supplemental Material \cite{sm}.

\subsection{\label{sec:level3}Generation rate and fidelity}
In order to increase the success probability of the generation protocol, we use the QMC to store a first photon while waiting for a second one. 
Knowing that a third photon detection is used to trigger the phase measurement sequence \cite{cotte_experimental_2022}, we can estimate the generation rate as a function of three parameters: the heralding rate of the single photons ($f_{\ket{1}}$), the success probability of the conditioning measurement, and a coefficient ($\beta_{\text{elec}}$) linked to electronic dead times (see Supplemental Material \cite{sm}).
\begin{equation}\label{taux}
f_{\ket{\psi}_{\text{cat}}}=\frac{f_{\ket{1}}}{3}\times \beta_{\text{elec}}\times P_{\text{success}}
\end{equation}
with:
\begin{equation}\label{success}
P_{\text{success}}=P(X\in[-\epsilon,\epsilon])\times P(\ket{1}\in[n_{min},n_{max}])
\end{equation}

\noindent
$P(\ket{1}\in[n_{min},n_{max}])$ being the probability to generate a single photon when the first photon has been stored between $n_{\min}$ and $n_{\max}$ round trips inside the QMC.

The counterpart to the increase in the generation rate via single-photon storage is a decrease in the fidelity of the generated cat state. The longer the first photon is stored inside the QMC, the lowest its fidelity is, resulting in an overall decrease in the produced cat state's fidelity. 
 
Two other parameters can be used to increase the SCS generation rate via an increase of the single photon generation rate. The first one is the selectivity of the spectral filter. If the spectral filter is less selective, the heralding rate will be higher. Consequently, the photon detection will herald a single photon that will be slightly multimode, reducing the mode matching between the single photons and the local oscillator. The second one is the pump power in the OPA.

By increasing the pump power, the squeezing factor of the OPA will increase, increasing the heralding rate of the single photons. In this case, the two-photon contribution in the squeezed vacuum will also increase, reducing the purity of the generated single photon. We show however, in the Supplemental Material \cite{sm}, that a loss of a few percent in the purity of the single photons does not have a tremendous impact on the fidelity of the generated state.

\subsection{\label{sec:level2}Experimental results}
In this section, we present the results we obtain by applying the aforementioned protocol with a single-photon heralding rate of \SI{310}{kHz}. 

\begin{figure}[h!]
\flushleft
 \hspace*{-0.2cm}
\includegraphics[width=0.5\textwidth]{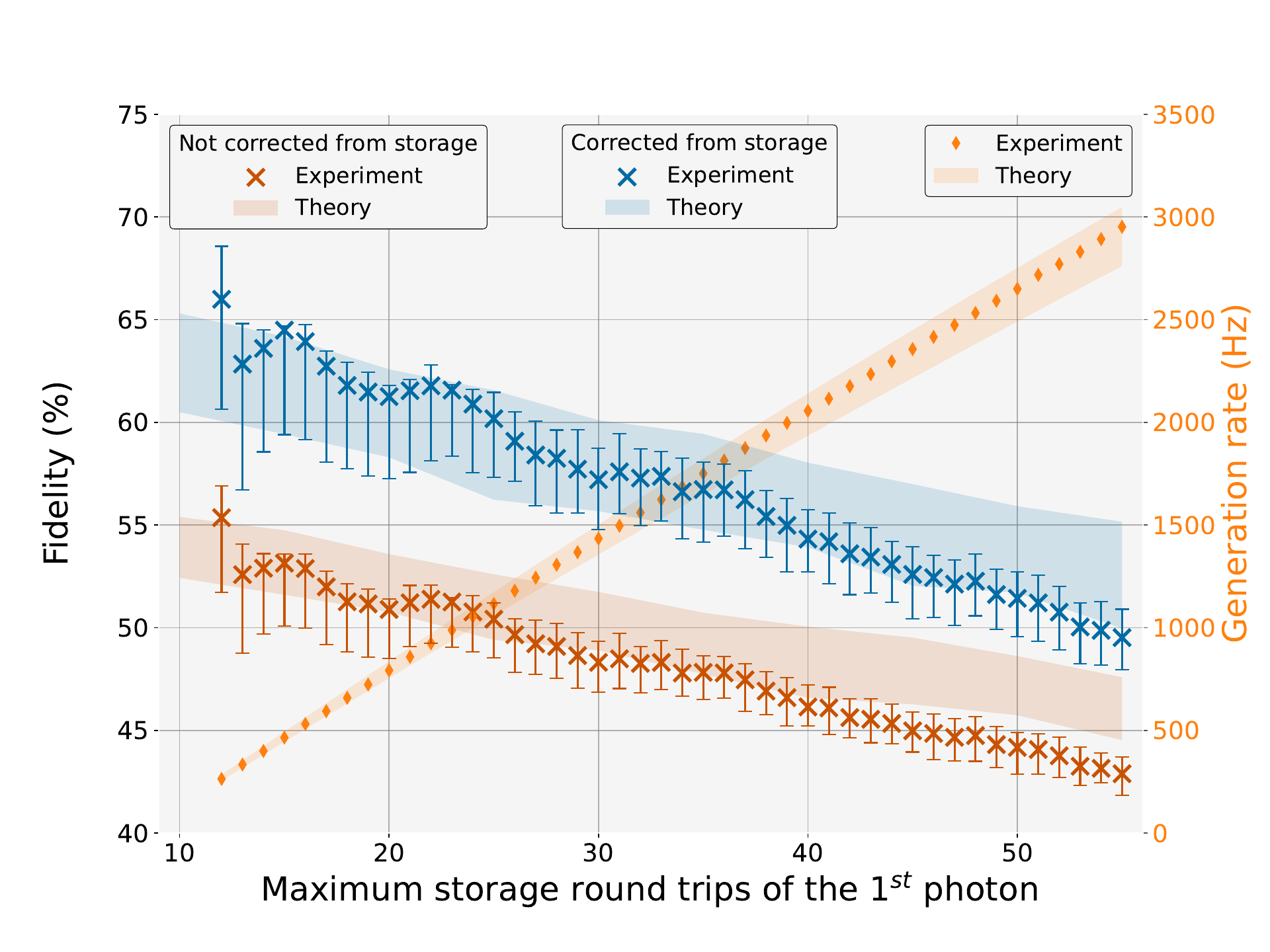}

\caption{\label{fig:fid_sto}\normalsize{Evolution of the generation rate (right axis) and of the fidelity (left axis) as a function of the maximum storage round trips of the $1^{\text{st}}$ photon. Experiment and theory with and without 15 round trips storage correction. See the Supplemental Material \cite{sm} for more information on the theory. The fidelity is corrected for a \SI{76}{\%} detection efficiency \cite{lvovsky_iterative_2004}. 
}}
\end{figure}
\begin{figure}[b!]
\vspace*{-1.5cm}
\includegraphics[width=\textwidth]{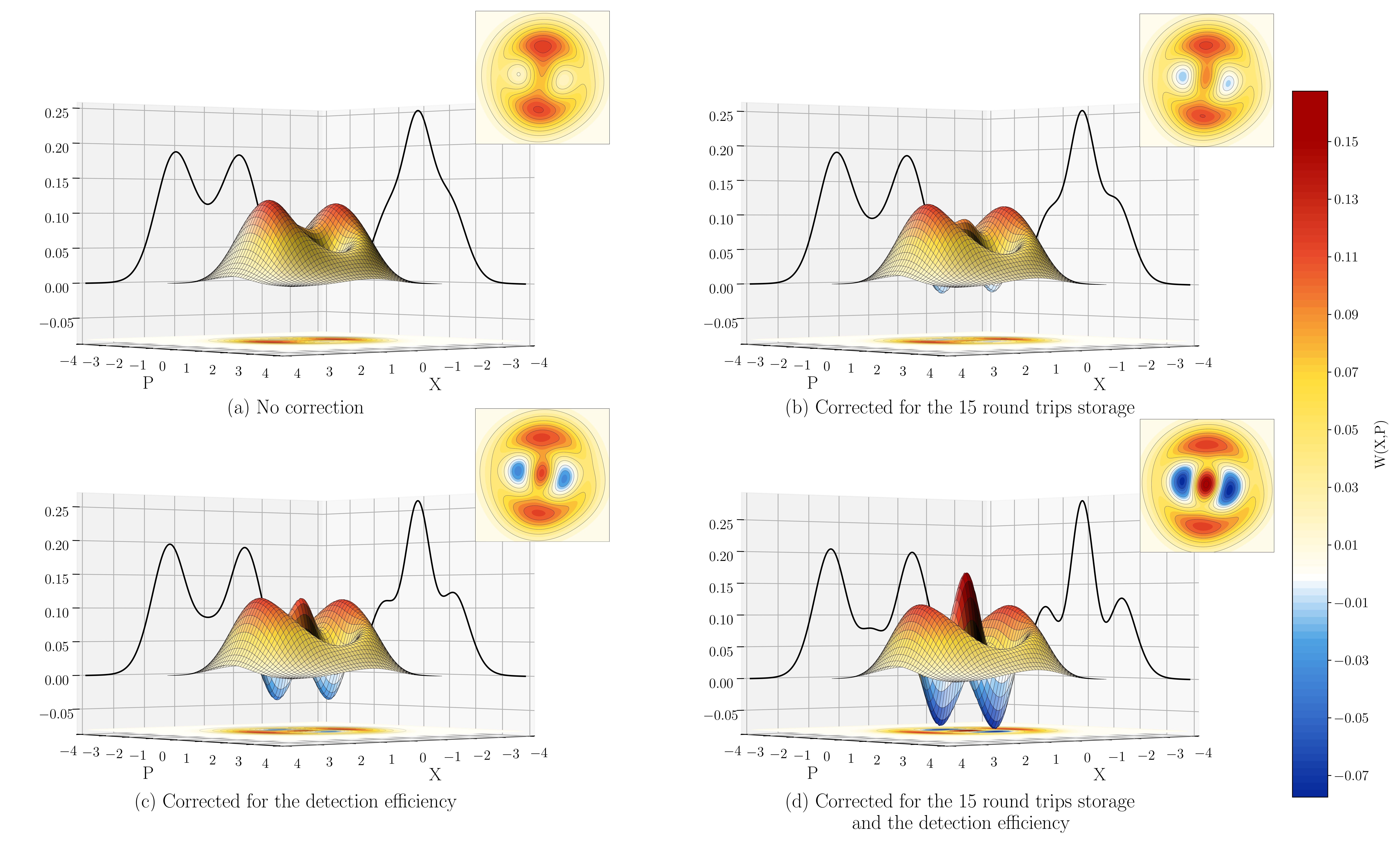}
\caption{\label{Wig} 
\normalsize{Wigner function of cat states generated at a rate of \SI{1}{kHz} stored 15 round trips: without corrections (a), corrected for the losses corresponding to a 15 round trip storage (b), stored 15 round trips corrected for the detection efficiency (c), corrected for the losses corresponding to a 15 round trip storage and for efficiency detection (d).}}
\label{fig:figures}
\vspace*{0.5cm}
\end{figure}

We characterize the produced states by performing quantum tomography using a MaxLike algorithm \cite{lvovsky_iterative_2004}.
For the data presented in Fig. \ref{fig:figures}, approximately 17000 measurements of SCS were used.
After a 15 round trips storage, the Wigner function presents two clear negative parts (-0.036 and -0.034) (see Fig.~\ref{fig:figures}) and a fidelity higher than \SI{51}{\%}, after correction for the $76\%$ detection efficiency.
To characterize the performances of our setup as a SCS source, we compute the fidelity of the SCS just after its generation (i.e., before the 15 round trips storage corresponding to a $15.9\%$ optical loss, see Supplemental Material \cite{sm}). The generated SCS has a fidelity to the target state $\ket{\psi}_{cat}$ higher than $60\%$.
We estimate the Wigner function of the state before the 15 round trips storage via the quantum tomography protocol by including a beam-splitter that accounts for storage losses in the model of the detection efficiency. By measuring the first
\cleardoublepage
\onecolumngrid

\twocolumngrid
\noindent photon's storage time for each SCS generation, we can post-select a subset of measurements for which the photon's storage time is lower than an arbitrary threshold. The reconstructed states can be understood as a statistical mixture of cat states of different fidelities. Fig. \ref{fig:fid_sto} presents the generation rate and the fidelity of the generated states as a function of the maximum number of storage round trips of the first photon inside the QMC.
The slightly non-monotonous evolution of the fidelity of the reconstructed Wigner functions can be explained by fluctuations of the mode matching between the stored single-photon and the local oscillator. This effect shows that the optical mode of the single photons does not exactly match an eigenmode of the QMC. This mode matching depends on fine alignments that can fluctuate in a week's time. In this case, the mode mismatch leads to a decrease of the fidelity that is slightly worse than what one could expect when considering only the optical losses of the QMC \cite{bouillard_quantum_2019}. This effect is not taken into account in the theoretical model plotted in Fig. \ref{fig:fid_sto} (see Supplemental Material \cite{sm}) and can explain the mismatch between the experimental data and the theoretical model for long storage time of the first photon.
We use a bootstrapping method \cite{efron_introduction_1994} to estimate the uncertainties associated with the quantum tomography, as detailed in the Supplemental Material \cite{sm}. 
\section{Conclusion}

In this letter, we present a proof of principle for a versatile scheme able to store and manipulate photons in real time.
We illustrate the performances of this setup with the generation of even Schrödinger cat states with a high fidelity (\SI{60}{\%}) and at a rate which is about one order of magnitude beyond the state of the art for such states on free-propagating light pulses (\SI{>1}{kHz}). The Wigner function of the generated SCS still shows clear negativity and a fidelity of more than \SI{50}{\%} after 15 storage round trips (\SI{197}{ns}). This storage time could give us multiple opportunities to perform a new breeding operation on the generated state to increase the amplitude of the generated non-Gaussian state. We obtain these results thanks to the real time control of the polarization state of 
the light pulse stored inside a low-loss quantum memory cavity. This ability reduces the impact of the non-deterministic nature of the breading protocols, thus offering promising perspectives for the development of more complex iterative schemes. The tools presented in this paper paves the way toward the generation of more complex non-Gaussian states without having to duplicate the experimental setup, as it is sometimes suggested in iterative protocols \cite{konno_logical_2024,sychev_enlargement_2017}. It is to be noticed that our setup uses relatively low efficiency ($\eta=0.45$) avalanche photodiodes. Since the overall generation rate of an $n$-step protocol is proportional to ($\eta^n$), we could expect an increase of a factor 4 in the generation rate by replacing our single photon detector with a $\eta=0.9$ detection efficiency superconducting nanowire single-photon detectors.

\begin{acknowledgments}
The authors thank Thorald Bergmann from Bergmann Messgeraete Entwicklung KG for his help with the control of the PCI BMESG08p. This work was supported by the Agence Nationale de la Recherche with the IGNITION project (Grant No. ANR-21-CE47-0015-01) and the NISQ2LSQ PEPR project (Grant No. ANR-22-PETQ-0006).
\end{acknowledgments}

\bibliography{biblio.bib}
\bibliographystyle{apsrev4-1}

\end{document}